\documentclass[12pt]{article}
\usepackage{graphicx}
\usepackage[bookmarksnumbered,colorlinks,plainpages]{hyperref}

\begin{document}

\title{Maximum path information and the principle of least action for chaotic system}

\author{Q.A. Wang\\
{\it Institut Sup\'erieur des Mat\'eriaux et M\'ecaniques Avanc\'es}, \\
{\it 44, Avenue F.A. Bartholdi, 72000 Le Mans, France}}

\date{}

\maketitle

\begin{abstract}
A path information is defined in connection with the different possible paths of
chaotic system moving in its phase space between two cells. On the basis of the
assumption that the paths are differentiated by their actions, we show that the
maximum path information leads to a path probability distribution as a function of
action from which the well known transition probability of Brownian motion can be
easily derived. An interesting result is that the most probable paths are just the
paths of least action. This suggests that the principle of least action, in a
probabilistic situation, is equivalent to the principle of maximization of information
or uncertainty associated with the probability distribution.
\end{abstract}


\vspace{1cm}

\section{Introduction}
The aim of this work is to investigate the probability distributions attributed to the
different paths of chaotic system moving between two points in the phase space
$\Gamma$. As usual, the phase space $\Gamma$ of a system is defined such that a point
in it represents a state of the system. For a $N$-body system moving in three
dimensional ordinary configuration space, the $\Gamma$-space is of $6N$ dimension
($3N$ positions and $3N$ momenta) if it can be smoothly occupied.

Now we look at a nonequilibrium system moving in the $\Gamma$-space between two
points, $a$ and $b$, which are in two elementary cells of a given partition of the
phase space (Figure 1). We will use the concept of trajectory or path of classical
mechanics. If the motion of the system is regular, or if the phase manyfold has
positive or zero riemannian curvature, there will be only one possible trajectory
between the two points, or, in other words, there will be only a fine bundle of paths
which track each other between the initial and the final cells. These trajectories
must be the paths minimizing action according to the principle of least
action\cite{Biesiada,Arnold} and have unitary probability. Any other path must have
zero probability.

For a system in chaotic motion, or when the riemannian curvature of the phase manyfold
is negative, the things are different\cite{Arnold}. Two points indistinguishable in
the initial cell can separate from each other exponentially. Normally, these two
points, after their departure from the initial cell, will never meet each other in a
final cell in the phase space. However, it is possible that they pass through a same
cell at two different times. So between two given phase cells represented in Figure 1
by $a$ and $b$, respectively, there may be many possible paths labelled by $k$
($k$=1,2,...w) with different travelling time $t_{ab}(k)$ of the system and different
probability $p_{ab}(k)$ for the system to take the path $k$. This is called path
probability distribution.

\begin{figure}[t] \label{f1}
\includegraphics[width=5in,height=6in]{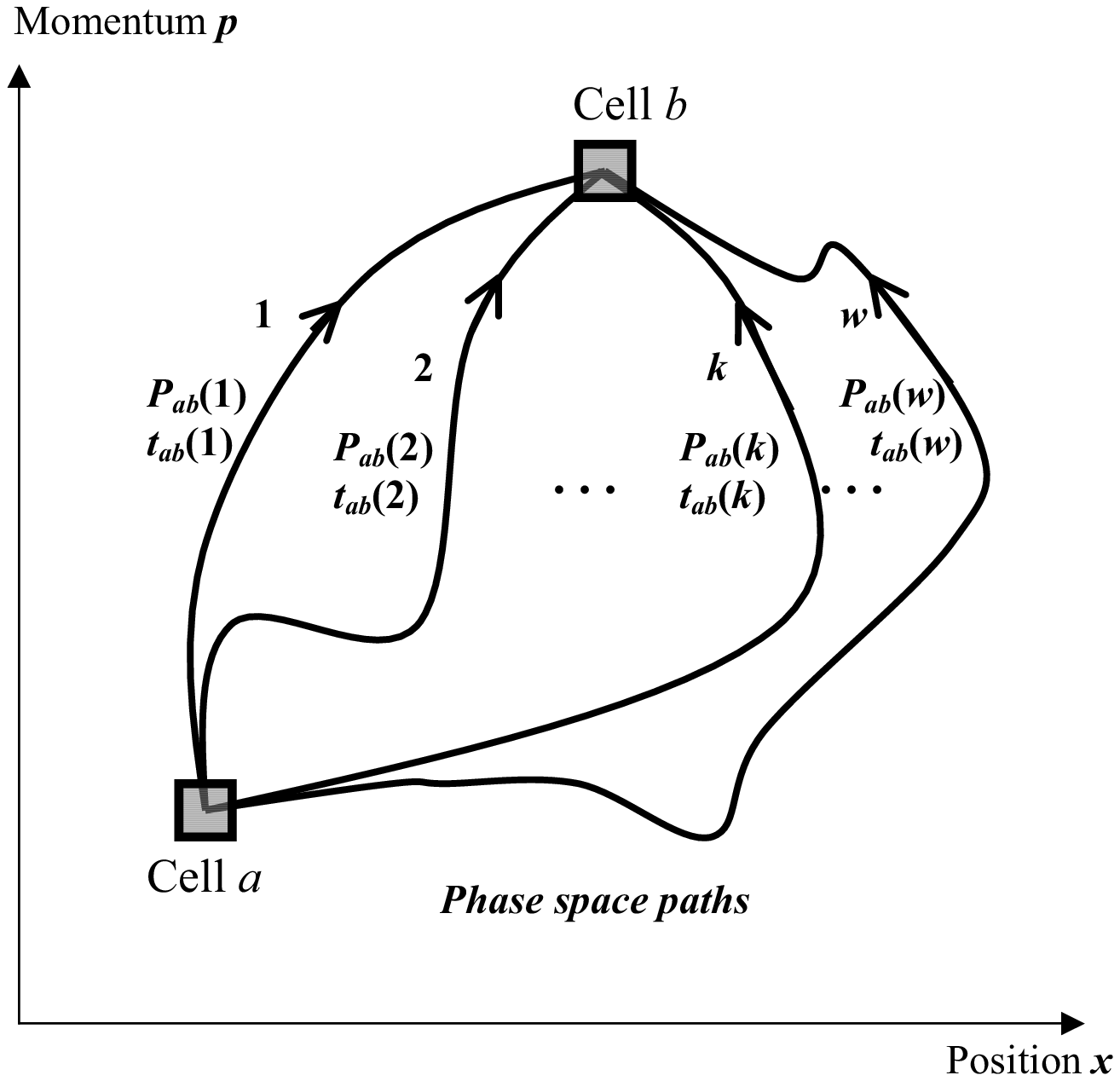}
\caption{Possible phase space paths ($k=1,2...w$) for a chaotic system to go from
the points in the phase cell $a$ to the points in the phase cell $b$ during the time
$t_{ab}(k)$ (along the path $k$).}
\end{figure}

The path probability distribution is defined as follows. Suppose an ensemble of a
large number $L$ of identical systems all moving in the phase space from cell $a$ to
cell $b$ with $w$ possible paths. We observe $L_k$ systems travelling on the path $k$
($k=1,2 ...w$). The probability $p_{ab}(k)$ that the system take the path $k$ is thus
defined as usual by $p_{ab}(k)=\frac{L_k}{L}$. We naturally have
$\sum_{k=1}^wp_{ab}(k)=1$. By definition, $p_{ab}(k)$ is a transition probability from
state $a$ to state $b$.

In this work, the path probability distribution due to dynamic instability is
studied in connection with information theory and the principle of least action.
First of all, we suppose that the different paths of the nonequilibrium systems
moving between the phase cells $a$ and $b$ are uniquely differentiated by their
action defined by
\begin{eqnarray}                                            \label{c5}
A_{ab}(k)=\int_{t_{ab}(k)}L_k(t)dt
\end{eqnarray}
where $L_k(t)$ is the Lagrangian of the system at time $t$ along the path $k$ and is
defined by $L_k(t)=U_k(t)-V_k(t)$ where $U_k(t)$ is the total kinetic energy and
$V_k(t)$ the total potential energy of the system. The integral is carried out along a
path $k$ during $t_{ab}(k)$, the travelling time of the system along the path. This is
an essential assumption of this work. If the paths can be identified only by their
actions, then it will be possible to study their probability distributions with the
information concept and the method of maximum information of Jaynes\cite{Jaynes} in
connection with our knowledge about action. This approach will leads us to a
probabilistic interpretation of the mechanical principle of Maupertuis and a
probability distribution depending on action. As an application, the obtained path
probability distribution will be used to derive the transition probability of Brownian
motion.

\section{A path information}
The information we address here is our ignorance about the system under consideration.
More we know about the system, less there is information. According to
Shannon\cite{Shannon}, this information can be measured by the formula
$S=-\sum_ip_i\ln p_i$ where $p_i$ is certain probability attributed to the situation
$i$. We usually ask $\sum_ip_i=1$ with a summation over all the possible situations.

Now for our ensemble of $w$ possible paths in Figure 1, a Shannon information can be
defined as follows :
\begin{eqnarray}                                            \label{c1}
H(a,b)=-\sum_{k=1}^wp_{ab}(k)\ln p_{ab}(k).
\end{eqnarray}
$H(a,b)$ is a {\it path information} and should be interpreted as the missing
information necessary for predicting which path a system of the ensemble takes from
$a$ to $b$.

According to our starting assumption, the quantity that differentiates the paths and
their probability of occurrence is the Lagrangian action. In what follows, a
statistics is developed on the basis of this assumption.

\section{Probability distribution of maximum information}

We consider the ensemble containing a large number of the studied system moving from
$a$ to $b$. These systems are distributed over the $w$ paths according to $p_{ab}(k)$
in connection with the action $A_{ab}(k)$. An expectation of the action over all the
possible paths can then be calculated by
\begin{eqnarray}                                            \label{c2}
A_{ab}=\sum_{k=1}^wp_{ab}(k)A_{ab}(k).
\end{eqnarray}
On the other hand, the path information $H(a,b)$ in Eq.(\ref{c1}) is concave as a
function of normalized probability $p_{ab}(k)$. According to the principle of Jaynes,
in order to get the optimal distribution, $H(a,b)$ can be maximized under the
constraints associated with our knowledge about the system and the relevant random
variables, i.e., with the normalization of $p_{ab}(k)$ and the expectation $A_{ab}$ :
\begin{eqnarray}                                            \label{c6x}
\delta\left[-H(a,b)+\alpha\sum_{k=1}^{w}p_{ab}(k)
+\eta\sum_{k=1}^{w}p_{ab}(k)A_{ab}(k)\right]=0
\end{eqnarray}
This leads to the following distribution
\begin{eqnarray}                                            \label{c6}
p_{ab}(k)=\frac{1}{Q}\exp[-\eta A_{ab}(k)]
\end{eqnarray}
Putting this distribution into $H(a,b)$ of Eq.(\ref{c1}), we get
\begin{eqnarray}                                            \label{c7}
H(a,b)=\ln Q+\eta A_{ab}=\ln Q-\eta \frac{\partial}{\partial\eta}\ln Q
\end{eqnarray}
where $Q$ is given by $Q=\sum_{k=1}^w\exp[-\eta A_{ab}(k)]$ and $A_{ab}$ is given
by
\begin{eqnarray}                                            \label{c7x}
A_{ab}=-\frac{\partial}{\partial\eta}\ln Q.
\end{eqnarray}

\section{Stability of the path probability distribution}
Now we shall show that the above distribution is stable with respect to the
fluctuation of action. Suppose that each path is cut into two parts $1$ (the segments
on the side of the cell $a$) and $2$ (the segments on the side of $b$). All the
segments 1 are contained in the group 1, and all the segments 2 in the group 2. Each
group has a path information $H_1=H_2=H$ and an average action $A_1=A_2=A$. The total
information is then $H(a,b)=H_1+H_2=2H$ and total average action is
$A(a,b)=A_1+A_2=2A$. If now we consider a small variation of the division of the paths
with virtual changes in the two groups such that $\delta A_1=\delta A=-\delta A_2$,
the total information will be changed and can be written as
\begin{eqnarray}                                            \label{8}
H'(a,b)=H(A+\delta A)+H(A-\delta A)
\end{eqnarray}
In view of the fact that the distribution Eq.(\ref{c6}) and the relationship
Eq.(\ref{c7}) are consequence of maximum information, the stability condition requires
that the information does not increase with the virtual changes of the two groups. We
must have
\begin{eqnarray}                                            \label{9}
\delta H=H'(a,b)-H(a,b)\leq 0,
\end{eqnarray}
i.e.,
\begin{eqnarray}                                            \label{10}
H(A+\delta A)+H(A-\delta A)-2H(A)\leq 0
\end{eqnarray}
which means
\begin{eqnarray}                                            \label{11}
\frac{\partial^2 H}{\partial A^2}\leq 0.
\end{eqnarray}
Now let us see if this stability condition is always fulfilled. From Eq.(\ref{c7}),
one gets $\frac{\partial^2 H}{\partial A^2}=\frac{\partial\eta}{\partial A}$. Then
considering the definition of average action Eq.(\ref{c2}), we straightforwardly
calculate
\begin{eqnarray}                                            \label{12}
\frac{\partial A}{\partial \eta}=-\sigma^2,
\end{eqnarray}
which implies
\begin{eqnarray}                                            \label{x12}
\frac{\partial^2 H}{\partial A^2}=-\frac{1}{\sigma^2}\leq 0
\end{eqnarray}
where the variance $\sigma^2=\overline{A^2}-\overline{A}^2\geq 0$ characterizes the
fluctuation of the action $A$. This proves the stability of the maximum information
distribution Eq.(\ref{c6}) with respect to the action fluctuation of the paths.

\section{Application to Brownian motion}
What is the parameter $\eta$ in the path probability distribution Eq.(\ref{c6})? A
possible physical meaning of $\eta$ can be found with a special example : Brownian
motion. Suppose a certain path in Figure 1 along which a Brownian particle moves from
$a$ to $b$ via an intermediate point or cell $k$. Between the three cells $a$, $k$,
and $b$, the particle is free. The action $A_{ab}(k)$ of the particle from $a$ to $b$
can be calculated to be\cite{Feynman}
\begin{eqnarray}                                            \label{x9}
A_{ab}(k)=\frac{m(x_k-x_a)^2}{2(t_k-t_a)}+\frac{m(x_b-x_k)^2}{2(t_b-t_k)}.
\end{eqnarray}
Then from Eq.(\ref{c6}), we have
\begin{eqnarray}                                            \label{c9}
p_{ab}(k) &=& \frac{1}{Q}\exp\left[-m\eta\frac{(x_k-x_a)^2}{2(t_k-t_a)}\right]
\exp\left[-m\eta\frac{(x_b-x_k)^2}{2(t_b-t_k)}\right]
\end{eqnarray}
On the other hand, it is known\cite{Kubo} that, as a solution of the diffusion
equation, the transition probability for the particle to go from $a$ to $b$ via $k$ is
\begin{eqnarray}                                            \label{c8}
p_{ab}(k) &=& p_{ak}p_{kb} \\\nonumber &=& \frac{1}{[4\pi
D(t_k-t_a)]^{d/2}}\exp\left[-\frac{(x_k-x_a)^2}{4D(t_k-t_a)}\right]\\\nonumber
&\times&\frac{1}{[4\pi
D(t_b-t_k)]^{d/2}}\exp\left[-\frac{(x_b-x_k)^2}{4D(t_b-t_k)}\right]
\end{eqnarray}
where $D$ is the diffusion coefficient supposed constant everywhere in the phase
space, $t_a$, $t_k$ and $t_b$ are the time and $x_a$, $x_k$ and $x_b$ the position
coordinates of the particle at $a$ $k$ and $b$, respectively, $d$ is the dimension of
the ordinary configuration space, $p_{ak}$ and $p_{kb}$ are the transition
probabilities of the particle from $a$ to $k$ and from $k$ to $b$, respectively. A
comparison of Eq.(\ref{c9}) and Eq.(\ref{c8}) leads to
\begin{eqnarray}                                            \label{xx9}
\eta=\frac{1}{2mD}
\end{eqnarray}
where $m$ is the mass of the particle. We see that the parameter $\eta$ is related
to the diffusion coefficient $D$.

For a system containing a large number of Brownian particles, the above result is
still valid. The only difference is that, in this case, there are just more
intermediate points and the total action will be calculated over all particles each
having a large number of intermediate points. The above result can be used for many
thermodynamic systems modelled with Brownian particles.

If we suppose that the self diffusion coefficient can be approximated by
Stokes-Einstein relation $D=\frac{k_BT}{6\pi \eta_0R}$ of a self-diffusion of the
Brownian particles in a dilute medium, we get
\begin{eqnarray}                                            \label{xx10}
\eta=\frac{6\pi \eta_0R}{2mk_BT}
\end{eqnarray}
where $\eta_0$ is the viscosity of the medium, $R$ is the radius of the Brownian
particles, $k_B$ is the Boltzmann constant and $T$ the temperature of the medium. In
this case, the path probability distribution Eq.(\ref{c6}) becomes
\begin{eqnarray}                                            \label{x6}
p_{ab}(k)=\frac{1}{Q}\exp[-\frac{\Lambda_{ab}(k)}{k_BT}].
\end{eqnarray}
where $\Lambda_{ab}(k)=\frac{6\pi\eta_0R}{2m} A_{ab}(k)$. This shows a temperature
dependence of path probability distribution.

\section{The principle of maximum information and the principle of least action}
Now we turn our attention to the connection between maximum path information and
least action. It can be shown (see below) that the paths of least action are the
most probable paths provided $\eta=\frac{\partial H(a,b)}{\partial
\overline{A}_{ab}}>0$. In fact, from Eq.(\ref{c6}), positive $\eta$ obviously
implies that the paths of smaller action are statistically more probable than the
paths of larger action. So the most probable paths must minimize the action.

This property of the distribution Eq.(\ref{c6}) can be mathematically discussed in
the same manner as the stability of the distribution proved in section 4. We still
consider the two groups $1$ and $2$ of path segments with $H_1=H_2=H$ and
$A_1=A_2=A$. The total information is then $H(a,b)=2H$ and the total average action
is $A(a,b)=2A$. Now suppose that the two groups are deformed such that $\delta
H_1=\delta H=-\delta H_2$. The total average action after the group deformation can
be written as
\begin{eqnarray}                                            \label{13}
A'(a,b) &=& A_1(H_1+\delta H_1)+A_2(H_2+\delta H_2)\\ \nonumber  &=& A(H+\delta
H)+A(H-\delta H).
\end{eqnarray}
If the distribution Eq.(\ref{c6}) and the relationship Eq.(\ref{c7}) correspond to
least action, the total average action after the group deformation can not decrease:
$\delta A=A'(a,b)-A(a,b)\geq 0$, i.e.,
\begin{eqnarray}                                            \label{14}
A(H+\delta H)+A(H-\delta H)-2A(H)\geq 0
\end{eqnarray}
which means
\begin{eqnarray}                                            \label{15}
\frac{\partial^2 A}{\partial H^2}\geq 0.
\end{eqnarray}
On the other hand, with the help of Eq.(\ref{c7}), we can prove:
\begin{eqnarray}                                            \label{16}
\frac{\partial^2 A}{\partial H^2}=-\frac{1}{\eta^2}\frac{\partial\eta}{\partial
H}=-\frac{1}{\eta^3}\frac{\partial\eta}{\partial A}.
\end{eqnarray}
Now considering $\frac{\partial\eta}{\partial A}=-\frac{1}{\sigma^2}$, we get
\begin{eqnarray}                                            \label{17}
\frac{\partial^2 A}{\partial H^2}=\frac{1}{\sigma^2\eta^3}.
\end{eqnarray}
We see that if Eq.(\ref{15}) is true, we have
\begin{eqnarray}                                            \label{18}
\eta\geq 0.
\end{eqnarray}
In other words, the positivity of $\eta$ implies that the principle of maximum path
information is intrinsically connected with the principle of least action: {\it the
most probable paths given by the distribution of maximum information are just the
paths of least action.}

In view of the relationship Eq.(\ref{xx9}), the positivity of $\eta$ is ensured by the
positivity of diffusion coefficient $D$. We can also invert this statement: if the
most probable paths derived from the probability distribution Eq.(\ref{c6}) minimize
the action, then the diffusion coefficient must be positive.

\section{Concluding remarks}
It is hoped that this work may contribute to the study of the behaviors of chaotic
systems. If there is no chaos, the path information will vanish and there will be
between two phase cells only a fine bundle of parallel paths which are the paths of
least action with unitary probability of occurrence. More the system under
consideration is chaotic, more there are possible paths with different actions, and
larger the information is. So we conjecture that the path information $H(a,b)$ may
be used as a measure of chaos, like Kolmogorov-Sinai entropy (KSE)\cite{Dorfman}. It
should be noted that there is an important difference between $H(a,b)$ and KSE.
$H(a,b)$ is an information associated with different paths relating two phase cells
but having arbitrarily different travelling time. In the language of discrete time
iteration, the different paths have arbitrarily different number of step of
iteration. On the other hand, KSE can be defined as an information associated with
different paths that leave the initial cell with arbitrary destination but same
travelling time, i.e., same number of step of iteration\cite{Dorfman}. Further
investigation is necessary in order to clarify the relationship between these two
information measures for chaotic systems.

The result of this work is in accordance with the path integral formulation of
quantum physics and the idea that dynamic systems can be viewed as geodesic flows on
the manifolds in $\Gamma$-space\cite{Biesiada,Arnold}. This idea was proposed
several decades ago and used in, among others, the description of violent relaxation
of self-gravitating $N$-body systems, in Yang-Mills theories and general
relativity\cite{Biesiada}. It is known that the least action solutions of the
Yang-Mills equation and the most probable least action geometries of spacetime
summed for Feynman path integrals correspond to instantons which are so important in
quantum field theory, quantum chromodynamics and cosmos evolution\cite{Naschie}. We
would like to mention here a recent development of gravitational instanton physics
which shows that a modification of Feynman path integrals in the fractal geometries
of $\varepsilon^{(\infty)}$ space (sum over dimensions instead of over paths) can
not only reinforce the physical interpretation of the existence of instantons, but
also allow one to calculate accurately the mass of several high energy elementary
particles\cite{Naschie}. This is an very encouraging result for the attempts to
calculate information, action integrals and Feynman path integrals in discontinuous
and nondifferentiable space (strange attractors, for example) in which the result of
present work has been used to derive the method of maximum entropy change for
dynamic systems moving in fractal phase space\cite{Wang04}.

Summarizing, a path information is defined for an ensemble of possible paths of
chaotic systems moving between two cells in phase space. It is shown that, if we
suppose that the different paths are physically identified by their actions, the
maximization of the path information leads to a path probability distribution as a
function of the action. As a special example, the transition probability of Brownian
particles is derived. In this case, we show that the most probable paths derived
from maximum information minimize the action. This suggests that the principle of
least action, in probabilistic situation, is equivalent to the principle of
maximization of information or uncertainty associated with the probability
distribution. This result may be considered as an argument supporting this method of
best guess\cite{Jaynes} for nonequilibrium systems.

\section*{Acknowledgments}

I acknowledge with great pleasure the useful discussions with Professors A. Le
M\'ehaut\'e, F. Tsobnang, L. Nivanen and M. Pezeril. Thanks are also due to
Professor El Naschie for valuable comments.

\end{document}